\documentclass[10pt,tightenlines,eqsecnum,floats,aps,amsmath,
amssymb,nofootinbib,prd,notitlepage]{revtex4-1}
\usepackage[utf8x]{inputenc}
\usepackage{MnSymbol}
\usepackage{tensor}
\newcommand{\QS}{Q_{\rm S}}
\newcommand{\QWZ}{Q_{\rm WZ}}
\newcommand{\LL}{\mathcal L}
\newcommand{\LLL}{\mathfrak L}
\newcommand{\iu}{\mathrm i}
\newcommand{\MM}{\mathcal{M}}
\newcommand{\SUU}{{\rm SU(3)/(U(1)}\times{\rm U(1))}}
\newcommand{\nnn}{\mathfrak{n}}
\newcommand{\nnno}{\mathfrak{n}_1}
\newcommand{\nn}{\mathbf{n}}
\newcommand{\mmm}{\mathfrak{m}}
\newcommand{\mmmo}{\mathfrak{m}_1}
\newcommand{\Tr}[1]{{\rm Tr}\left(#1\right)}

\newcommand{\kkt}{\alpha}
\newcommand{\kke}{\beta}
\newcommand{\vt}{v_\kkt}
\newcommand{\ve}{v_\kke}
\newcommand{\go}{g_1}
\newcommand{\gh}{g_{\frac{1}{2}}}
\DeclareMathOperator{\im}{im}
\begin{document}
\title{Integral expression for a topological charge in the Faddeev-Niemi non-linear sigma model}
\author{Marcin Kisielowski}
\email{Marcin.Kisielowski@fuw.edu.pl}
\affiliation{Instytut Fizyki Teoretycznej, Uniwersytet Warszawski,\\
ul. Ho{\.z}a 69, 00-681 Warszawa (Warsaw), Polska (Poland)}
\affiliation{St. Petersburg Department of Steklov Mathematical Institute,\\
Russian Academy of Sciences, Fontanka 27, St. Petersburg, Russia}
\begin{abstract}
We have introduced Faddeev-Niemi type variables for static SU(3) Yang-Mills theory. The variables suggest that a non-linear sigma model whose sigma fields take values in SU(3)/(U(1)xU(1)) and SU(3)/(SU(2)xU(1)) may be relevant to infrared limit of the theory. Shabanov showed that the energy functional of the non-linear sigma model is bounded from below by certain functional. However, Shabanov's functional is not homotopy invariant, and its value can be an arbitrary real number -- therefore it is not a topological charge. Since the third homotopy group of SU(3)/(U(1)xU(1)) is isomorphic to the group of integer numbers, there is a non-trivial topological charge (given by the isomorphism). We apply Novikov's procedure to obtain integral expression for this charge. The resulting formula is analogous to the Whitehead's realization of the Hopf invariant.
\end{abstract}
\maketitle
\section{Introduction}
Faddeev and Niemi introduced new variables for the SU(2) Yang-Mills theory \cite{Faddeev_Niemi_SU2}, which reveal a structure of Faddeev-Skyrme non-linear sigma model \cite{sigma} in the Lagrangian. The non-linear sigma model is expected to have topologically non-trivial excitations: in the static case, one assumes boundary conditions that compactify the domain to $\mathbb{S}^3$, and therefore the sigma field, being a map from $\mathbb{S}^3$ to $\mathbb{S}^2$, is characterized by the third homotopy group of the sphere $\pi_3(\mathbb{S}^2)$. The corresponding topological charge is given by the Hopf integral. It bounds from below the energy functional of the Skyrme-Faddeev model \cite{VK}. This property suggested, that there are topologically non-trivial solutions of the model. Numerical investigations \cite{num} showed that such solutions indeed may exist.

We proposed a Faddeev-Niemi type decomposition for the static SU(3) Yang-Mills theory that reveals a structure of a non-linear sigma model in the Lagrangian \cite{FNK}. The non-linear sigma model has been introduced by Faddeev and Niemi as a generalization of the Faddeev-Skyrme model that may be relevant to infrared limit of SU(3) Yang-Mills theory \cite{Faddeev_Niemi_SUN}. In this model the sigma fields take values in $\SUU$ and SU(3)/(SU(2)$\times$U(1)) \cite{Faddeev_Niemi_SUN}. Given a map $g:\mathbb{S}^3\to {\rm SU(3)}$, the sigma fields are
$$
\nnn=g\kappa_3 g^\dagger,\quad \mmm= g\kappa_8 g^\dagger,
$$
where $\kappa_a=-\frac{\iu}{\sqrt{2}}\lambda_a, a\in\{1,2,\ldots,8\}$ form a basis of su(3), $\lambda_a$ are the Gell-Mann matrices. On the orbits $\SUU$ and SU(3)/(SU(2)$\times$U(1)) there are natural symplectic forms, called the Kirillov-Kostant symplectic forms. The pull-backs of those forms via the maps $\nnn$ and $\mmm$, respectively, are:
$$
F=\frac{1}{4\sqrt{2}}(\nnn, [d\nnn,d\nnn]+[d\mmm,d\mmm]),\quad G=\frac{1}{4\sqrt{2}}(\mmm, [d\nnn,d\nnn]+[d\mmm,d\mmm]),
$$
where $(C,D)=\Tr{C^\dagger D}$ for $C,D\in \rm{su(3)}$, $[d\nnn,d\nnn]=d\nnn^a\wedge d\nnn^{b}[\kappa_a,\kappa_b],\, [d\mmm,d\mmm]=d\mmm^a\wedge d\mmm^{b}[\kappa_a,\kappa_b]$, $\nnn=\nnn^a\kappa_a,\,\mmm=\mmm^a \kappa_a$. The forms $F$ and $G$ are closed and since $H^2(\mathbb{S}^3)=0$ they are also exact, i.e. there exist one-forms $A,B \in \Omega^1(\mathbb{S}^3)$ such that $F=dA,\,G=dB$. Shabanov showed, that the energy functional of the non-linear sigma model is bounded from below by a functional $\QS$ \cite{Shabanov}:
$$
\QS[\nnn]=\frac{1}{4 \pi^2}\int_{\mathbb{S}^3} (A\wedge F + B\wedge G).
$$
Note, that we do not write the dependence on $\mmm$. It is because $\mmm$ is completely determined by $\nnn$. For more details on this issue, see the beginning of the section \ref{sc:charge_chiral_fields}.

The problem is that the functional $\QS$ is not homotopy invariant and its value can be an arbitrary real number. The problem is illustrated on the following examples. Let $U:\mathbb{S}^3\to \rm{SU(2)}$ be a map whose degree is 1, i.e.
$$
\frac{1}{24\pi^2}\int_{\mathbb{S}^3}\Tr{U^{-1}dU\wedge U^{-1}dU\wedge U^{-1}dU}=1.
$$
Consider a field $\go:\mathbb{S}^3\to \rm{SU(3)}$ defined by
$$
(\go)\indices{^i_j}=\frac{1}{2}\Tr{\sigma_i U \sigma_j U^\dagger},
$$
where $\sigma_i$ are the Pauli matrices.
It is easy to calculate, that the Wess-Zumino invariant for this map is:
$$
\QWZ[g]=\frac{1}{24\pi^2}\int_{\mathbb{S}^3}\Tr{g^{-1}dg\wedge g^{-1}dg\wedge g^{-1}dg}=4.
$$
It means in particular that the map $\go$ is not homotopic to a constant map. On the other hand 
$$
\QS[\nnno]=0,{\rm\ where\ } \nnno=\go\kappa_3 \go^\dagger\,(\mmmo=\go\kappa_8\go^\dagger).
$$
The Shabanov functional takes the value zero also on the constant map. From the long exact sequence for fibration (see e.g. \cite{BT}):
\begin{eqnarray*}
\ldots\to\pi_3({\rm U(1)}\times {\rm U(1)})\to \pi_3({\rm SU(3)}) \xrightarrow{\pi_*} \pi_3(\SUU)\to \pi_2({\rm U(1)}\times {\rm U(1)})\to \ldots \\ \ldots\to  \pi_0({\rm SU(3)}) \to \pi_0(\SUU)\to 0
\end{eqnarray*}
it is easy to infer, that the homotopy groups $\pi_3({\rm SU(3)})$ and $\pi_3(\SUU)$ are isomorphic, and the isomorphism is given by the projection $\pi:{\rm SU(3)}\to \SUU$. This means that $\nnno$ is not homotopic to a constant map, however the value of the Shabanov functional on both maps is equal 0. Therefore the Shabanov functional is not homotopy invariant. 

Moreover, a value of the Shabanov functional can be an arbitrary real number. To see this, consider a map $U':\mathbb{S}^3\to \rm{SU(2)}$ of degree 4:
$$
\frac{1}{24\pi^2}\int_{\mathbb{S}^3}\Tr{{U'}^{-1}dU'\wedge {U'}^{-1}dU'\wedge {U'}^{-1}d{U'}}=4.
$$
Consider also a map $\gh':\mathbb{S}^3\to \rm{SU(3)}$,
$$
\gh'=\left(\begin{array}{cc}
U'&0\\0&1
\end{array} \right).
$$
Let $\nnn'_{\frac{1}{2}}=\gh'\kappa_3(\gh')^\dagger\,(\mmm'_{\frac{1}{2}}=\gh'\kappa_8(\gh')^\dagger)$. It is easy to check, that $ \QWZ[\gh']=4$ as well as $\QS[\nnn'_{\frac{1}{2}}]=4$. Let us recall, that also $\QS[\nnno]=4$. Therefore there exists a homotopy $f_t:\mathbb{S}^3 \to \rm{SU(3)}$, $t\in [0,1]$ such that $f_0=\go,\,f_1=\gh'$. Calculating the corresponding Shabanov charges at each instant $t$ we obtain a continuous function $t\mapsto\QS^t$ such that $\QS^0=0,\, \QS^1=4$. Therefore for every $r\in [0,4]$ there exists $t_r\in [0,1]$ such that $\QS^{t_r}=r$. This shows that $\QS$ can take any value in the interval $[0,4]$. Obviously, the reasoning can be repeated for a map $U:\mathbb{S}^3\to \rm{SU(2)}$ with arbitrary degree and therefore the possible values of $\QS$ is the whole set of real numbers.

Our goal is to construct a functional $Q$, such that for each $g:\mathbb{S}^3\to{\rm SU(3)}$ it satisfies:
$$
Q[\pi(g)]=\QWZ[g],
$$
where $\QWZ[g]=\frac{1}{24\pi^2}\int_{\mathbb{S}^3}\Tr{g^{-1}dg\wedge g^{-1}dg\wedge g^{-1}dg}$ is the Wess-Zumino invariant. Such functional takes only integer values and defines an isomorphism of groups $\pi_3(\SUU)$ and $\mathbb{Z}$. We obtain this goal by using Novikov's procedure \cite{Novikov}.
\section{The cohomology ring of SU(3)/(U(1)$\times$U(1))}
A preparatory step in Novikov's procedure \cite{Novikov} is to calculate the cohomology ring of $\SUU$. In this section we find the cohomology ring of $\SUU$ using the theorem 22.1 of the seminal paper by Chevalley and Eilenberg on cohomology of Lie groups \cite{CE}.

Let us fix some notation: let $\Omega^p=\bigwedge^p {\rm su(3)}^*$ be the space of skew-symmetric linear functions on su(3), and $\Omega=\bigoplus_p \Omega^p$. From theorem 13.1. of \cite{CE} follows, that the invariant differential k-forms on SU(3)/(U(1)$\times$U(1)) are in 1-1 correspondence with the forms $\omega\in\Omega^p$, such that:
\begin{enumerate}
\item $\omega(X_1,\ldots, X_p)=0$ if at least one $X_i$ lies in u(1)$\oplus$u(1),
\item $\omega([X_1,X],X_2,\ldots, X_p)+\ldots+\omega(X_1,X_2,\ldots, [X_p,X])=0$ for $X\in {\rm u(1)}\oplus {\rm u(1)}$, $X_1,\ldots, X_p\in {\rm su(3)}$.
\end{enumerate}
A form $\omega\in \Omega^p$ satisfying those conditions is called orthogonal to u(1)$\oplus$u(1). Denote by $\Omega^p_{\rm ort}$ the subspace of such forms. We denote by $d:\Omega^p\to \Omega^{p+1}$ a differential
$$
(d\omega) (X_1,\ldots,X_{p+1})=\frac{1}{p+1}\sum_{k<l} (-1)^{k+l} \omega([X_k,X_l],X_1,\ldots,\hat{X_k},\ldots, \hat{X_l},\ldots, X_{p+1}).
$$
Let $Z^p=\ker d:\Omega^p\to \Omega^{p+1}$ be the cocycles of dimension $p$, and $B^p=\im d:\Omega^{p-1}\to\Omega^p$ the coboundaries of dimension $p$. Let $Z^p_{\rm ort}=Z^p\cap \Omega^p_{\rm ort}$ and $B^p_{\rm ort}=d \Omega^{p-1}_{\rm ort}$. The relative cohomology group of su(3) mod u(1)$\oplus$u(1) is defined as the quotient space $H^q_{\rm ort}=Z^p_{\rm ort}/ B^p_{\rm ort}$. From theorem 22.1 of \cite{CE} now follows that the cohomology groups $H^p_{\rm ort}$ and the p-th de Rham cohomology group of $\SUU$, $H^p_{\rm dR}\left({\rm SU(3)/(U(1)}\times{\rm U(1))}\right)$, are isomorphic. We will now calculate explicitly $H^p_{\rm ort}$.

Since $\Omega^1_{\rm ort}=0$, it immediately follows that $H^1_{\rm ort}=0$. We look now for $H^2_{\rm ort}$.  Denote by $\LL^a$ the basis dual to $\kappa_a=-\frac{\iu}{\sqrt{2}}\lambda_a$: $\LL^a(\kappa_b)=\delta^{a}_b$. It will be convenient for us to use the Cartan-Weyl basis:
$$
 \kappa_1^{\pm}:=\frac{1}{\sqrt{2}}(\kappa_1\pm \iu\kappa_2),\quad \kappa_2^{\pm}:=\frac{1}{\sqrt{2}}(\kappa_4\mp \iu\kappa_5),\quad \kappa_3^{\pm}:=\frac{1}{\sqrt{2}}(\kappa_6\pm \iu\kappa_7),\quad \kappa_3,\quad \kappa_8.
$$
The dual basis will be denoted analogously by: $\LL^\pm_{j},\ j\in\{1,2,3\}$, $\LL^3$, $\LL^8$. Let us underline, that although we use complex basis, we consider cohomology with real coefficients, i.e. we require that the forms considered are real-valued. Any two-form $\omega\in \Omega^2_{\rm ort}$ is of the following form:
$$
\omega= \omega_1\, \iu \LL_1^+\wedge \LL_1^-+\omega_2\, \iu \LL_2^+\wedge \LL_2^- + \omega_3\,\iu \LL_3^+\wedge \LL_3^-,
$$
where $\omega_1,\,\omega_2,\,\omega_3\in \mathbb{R}$. The cocycle condition $d \omega = 0$ leads to a condition on $\omega_1, \omega_2, \omega_3$:
$$
\omega_1+\omega_2+\omega_3=0.
$$
We choose a basis in the cocycle space $Z^2_{\rm ort}$:
\begin{equation}\label{eq:basis}
\kkt= -\iu \LL_1^+\wedge \LL_1^-+ \frac{\iu}{2}\LL_2^+\wedge \LL_2^-+\frac{\iu}{2}\LL_3^+\wedge \LL_3^-,\quad
\kke=\frac{\sqrt{3}}{2} \iu \LL_2^+\wedge \LL_2^- -\frac{\sqrt{3}}{2}\iu \LL_3^+\wedge \LL_3^-.
\end{equation}
The corresponding forms on SU(3)/(U(1)$\times$U(1)) and SU(3)/(SU(2)$\times$U(1)) are the Kirillov-Kostant symplectic forms. Since $B^2_{\rm ort}=0$, it follows that $H^2_{\rm ort}= Z^2_{\rm ort}=\mathbb{R}\times\mathbb{R}$.

The space $\Omega^3_{\rm ort}$ is 2-dimensional. The basis elements are:
$$
\rho=\frac{1}{2}\LL_1^+\wedge \LL_2^+ \wedge \LL_3^+ + \frac{1}{2}\LL_1^-\wedge \LL_2^- \wedge \LL_3^- {\rm \ and\ } \sigma=\frac{\iu}{2} \LL_1^+\wedge \LL_2^+ \wedge \LL_3^+ -\frac{\iu}{2} \LL_1^-\wedge \LL_2^- \wedge \LL_3^-.
$$
The form $\rho$ is exact $\rho=d(-\frac{\iu}{2} \LL_1^+\wedge \LL_1^-)$ and the form $\sigma$ is not closed:
$$
d \sigma = \LL_1^+\wedge \LL_1^-\wedge \LL_2^+\wedge \LL_2^- + \LL_1^+\wedge \LL_1^-\wedge \LL_3^+\wedge \LL_3^- +\LL_2^+\wedge \LL_2^-\wedge \LL_3^+\wedge \LL_3^-.
$$ Therefore $H^3_{\rm ort}=0$.

A 4-form $\omega\in \Omega^4_{\rm ort}$, can be written in the following way:
$$
\omega = \omega_{12}\, \LL_1^+\wedge \LL_1^-\wedge \LL_2^+\wedge \LL_2^- + \omega_{23}\, \LL_2^+\wedge \LL_2^-\wedge \LL_3^+\wedge \LL_3^- + \omega_{31}\, \LL_3^+\wedge \LL_3^-\wedge \LL_1^+\wedge \LL_1^-,
$$
where $\omega_{12},\,\omega_{23},\,\omega_{31}\in \mathbb{R}$. It is easy to verify, that $d \omega =0$ for any $\omega\in \Omega^4_{\rm ort}$. Therefore $Z^4_{\rm ort}=\Omega^4_{\rm ort}$, and the basis of $Z^4_{\rm ort}$ is:
\begin{eqnarray*}
&\kkt\wedge \kkt = \LL_1^+\wedge \LL_1^-\wedge \LL_2^+\wedge \LL_2^- + \LL_1^+\wedge \LL_1^-\wedge \LL_3^+\wedge \LL_3^- - \frac{1}{2}\LL_2^+\wedge \LL_2^-\wedge \LL_3^+\wedge \LL_3^-,\\
&\kke\wedge \kke = \frac{3}{2}\LL_2^+\wedge \LL_2^-\wedge \LL_3^+\wedge \LL_3^-,\\
&\kkt\wedge \kke= \frac{\sqrt{3}}{2} \LL_1^+\wedge \LL_1^-\wedge \LL_2^+\wedge \LL_2^- - \frac{\sqrt{3}}{2} \LL_1^+\wedge \LL_1^-\wedge \LL_3^+\wedge \LL_3^-.
\end{eqnarray*}
In particular $d \sigma$ can be expressed in this basis giving: 
$$
d\sigma= \kkt\wedge \kkt + \kke\wedge \kke.
$$
The space $B^4_{\rm ort}$ is 1-dimensional, and spanned by $\kkt\wedge \kkt+\kke\wedge \kke$. Therefore $H^4_{\rm ort}=\mathbb{R}\times\mathbb{R}$.

Since $\Omega^5_{\rm ort}=0$, it follows that $H^5_{\rm ort}=0$. The space $\Omega^6_{\rm ort}$ is one-dimensional, spanned by:
$$
\kkt\wedge \kkt \wedge \kkt= \LL_1^+\wedge \LL_1^-\wedge \LL_2^+\wedge \LL_2^-\wedge \LL_3^+\wedge \LL_3^-.
$$
Therefore $H^6_{\rm ort}= Z^6_{\rm ort}= \Omega^6_{\rm ort}=\mathbb{R}$.

As a result the cohomology ring $H_{\rm ort}^*:=\bigoplus_{p=1}^6 H_{\rm ort}^p$ is generated by $[\kkt]$ and $[\kke]$ satisfying:
$$
[\kkt]\wedge [\kkt]+[\kke]\wedge[\kke] = [d \sigma]=[0], \quad [\kke]\wedge[\kke]\wedge[\kke]=[0].
$$
The cohomology ring can be also inferred from the fact, that SU(3)/(U(1)$\times$U(1)) is a flag manifold \cite{BT}:
$$
H_{\rm dR}({\rm SU(3)/(U(1)}\times{\rm U(1))})=\mathbb{R}\left[[x_1],[x_2],[x_3]\right]\big/\left((1+[x_1])(1+[x_2])(1+[x_3])=1\right).
$$
This coincides with the ring we found under the identification:
\begin{equation}
x_1= \frac{1}{2\pi}\left(\kkt+\frac{\sqrt{3}}{3}\kke \right),\quad x_2= \frac{1}{2\pi}\left(-\kkt+\frac{\sqrt{3}}{3}\kke \right),\quad x_3=-\frac{1}{2\pi} \frac{2\sqrt{3}}{3}\kke.
\end{equation}
\section{The integral expression for the topological charge}
Having found the cohomology ring of $\SUU$ we can now construct the minimal model of $\SUU$ and apply Novikov's procedure \cite{Novikov} to obtain the integral expression for the topological charge.
\subsection{Sullivan's minimal model}
Let us recall the definition of a minimal model of a manifold $M$ \cite{BT}. Let $\MM=\bigoplus_{p\geq 0} \MM^p$ be a differential graded algebra, and $\Omega^*(M)$ the algebra of differential forms on $M$. We say that an element in $\MM$ is decomposable if it is a sum of products of positive elements in $\MM$, i.e. $m\in \MM^+\cdot \MM^+$, where $\MM^+=\bigoplus_{p>0}\MM^p$. A differential graded algebra $\MM$ is called a minimal model of a manifold $M$ if:
\begin{itemize}
\item $\MM$ is free;
\item there is a chain map $\Psi:\MM\to \Omega^*(M)$ which induces an isomorphism in cohomology;
\item the differential of a generator is either zero or decomposable.
\end{itemize}
The minimal model $\MM$ of the manifold SU(3)/(U(1)$\times$U(1)) has:
\begin{itemize}
\item two generators in dimension 2: $\widetilde{\kkt}, \widetilde{\kke}$,
\item one generator in dimension 3: $ \widetilde{\sigma}$,
\item one generator in dimension 5: $\tau$.
\end{itemize}
The generators satisfy:
$$
d\widetilde{\sigma}=(\widetilde{\kkt})^2+(\widetilde{\kke})^2,\quad d\tau = (\widetilde{\kke})^3.
$$
Let $\overline{\kkt},\ \overline{\kke}, \overline{\sigma}$ denote the differential forms on $\SUU$ corresponding to $\kkt,\, \kke, \sigma$. The map $\Psi:\MM\to \Omega^*({\rm SU(3)/(U(1)}\times{\rm U(1)})$ is given by:
$$
\widetilde{\kkt}\mapsto \overline{\kkt},\quad \widetilde{\kke}\mapsto \overline{\kke},\quad \widetilde{\sigma}\mapsto \overline{\sigma},\quad \tau\mapsto 0.
$$
\subsection{Novikov's construction}
Since $\dim \SUU > \dim \mathbb{S}^3$, the Novikov's construction reduces to the following procedure. Consider an extension $\overline{C}^q(\MM)$ of $\MM$ obtained by adding new free generators $\vt,\, \ve \in \overline{C}^q(\MM)$ whose $d$-operator lies in $\MM$:
$$
d\vt=\widetilde{\kkt},\quad d\ve=\widetilde{\kke}.$$
Let $\widetilde{z}=\vt\,\widetilde{\kkt}+\ve\,\widetilde{\kke}-\widetilde{\sigma}$. Clearly $\widetilde{z}$ is a cocycle and therefore represents a cohomology class $[\widetilde{z}]\in H^3(\overline{C}^q(\MM))$.

Consider a map $\nn:\mathbb{S}^3\to\SUU$. The map induces a homomorphism:
$$
\nn^* \psi : \MM \to \Omega^*(\mathbb{S}^3).
$$
Since $H^q_{\rm dR}(\mathbb{S}^3)=0$ for $q=1,2$, the homomorphism $\nn^*\psi$ extends to a homomorphism
$$
\psi_\nn: \overline{C}^q(\MM)\to \Omega^*(\mathbb{S}^3),
$$
called geometric realization. The extension is not unique, however the resulting topological charge does not depend on the choice of geometric realization. We denote images of $\vt,\,\ve,\,\widetilde{\kkt},\widetilde{\kke},\widetilde{z}$ under $\psi_\nn$ by $A,B,F,G,Z$. The topological charge is proportional to the integral:
$$
Q[\nn]\propto\int_{\mathbb{S}^3}( A\wedge F + B\wedge G - Z ).
$$
The proportionality factor will be calculated in the next subsection by the requirement, that for each $g:\mathbb{S}^3\to{\rm SU(3)}$ it satisfies:
$$
Q[\pi(g)]=\frac{1}{24\pi^2}\int_{\mathbb{S}^3} \Tr{g^{-1}dg\wedge g^{-1}dg \wedge g^{-1}dg}.$$
\subsection{The topological charge}
In order to fix the proportionality factor, let us first re-write the Wess-Zumino invariant using the basis $\kappa_j^\pm,\,\kappa_3,\,\kappa_8$ of su(3):
$$
\QWZ[g]=\frac{1}{24 \pi^2} \int_{\mathbb{S}^3}\left( \LLL^3\wedge d \LLL^3 + \LLL^8\wedge d \LLL^8 + \LLL^+_1\wedge d \LLL^-_1+\LLL^-_1\wedge d \LLL^+_1+\LLL^+_2\wedge d \LLL^-_2+\LLL^-_2\wedge d \LLL^+_2+\LLL^+_3\wedge d \LLL^-_3+\LLL^-_3\wedge d \LLL^+_3\right),
$$
where $\LLL^3:=\Tr{(\kappa_3)^\dagger g^{-1}dg},\,\LLL^8:=\Tr{(\kappa_8)^\dagger g^{-1}dg},\,\LLL^\pm_j:=\Tr{(\kappa^\pm_j)^\dagger g^{-1}dg},\,j=1,2,3$. The proportionality factor is fixed using the following equality:
$$
\QWZ[g]=\frac{1}{8 \pi^2} \int_{\mathbb{S}^3}\left( \LLL^3\wedge d\LLL^3+\LLL^8\wedge d\LLL^8-\iu \LLL^+_1\wedge \LLL^+_2\wedge \LLL^+_3 +\iu \LLL^-_1\wedge \LLL^-_2\wedge \LLL^-_3\right)=\frac{1}{4\pi^2} \int_{\mathbb{S}^3}\left( A\wedge F + B\wedge G - Z\right).
$$
Therefore we define the topological charge to be:
$$
Q[\nn]=\frac{1}{4\pi^2} \int_{\mathbb{S}^3} \left( A\wedge F + B\wedge G - Z \right),
$$
where $\nn$ is a map from $\mathbb{S}^3$ to $\SUU$,
$$
F= -\iu \LLL_1^+\wedge \LLL_1^-+ \frac{\iu}{2}\LLL_2^+\wedge \LLL_2^-+\frac{\iu}{2}\LLL_3^+\wedge \LLL_3^-,\quad
G=\frac{\sqrt{3}}{2} \iu \LLL_2^+\wedge \LLL_2^- -\frac{\sqrt{3}}{2}\iu \LLL_3^+\wedge \LLL_3^-
$$
are pull-backs of the forms $\overline{\alpha},\,\overline{\beta}$ with the map $\nn$, $dA = F,\, dB=G$ and
$$
Z=\frac{\iu}{2} \LLL_1^+\wedge \LLL_2^+ \wedge \LLL_3^+ -\frac{\iu}{2} \LLL_1^-\wedge \LLL_2^- \wedge \LLL_3^-
$$
is the pull-back of the form $\overline{\sigma}$ with the map $\nn$.
\section{Topological charge in terms of sigma fields}\label{sc:charge_chiral_fields}
The field variables in the Faddev-Niemi non-linear sigma model \cite{Faddeev_Niemi_SUN, Shabanov,FNK} are two sigma fields:
$$
\nnn=g \kappa_3 g^\dagger ,\quad \mmm= g \kappa_8 g^\dagger.
$$
It is therefore instructive to express the forms $F,G,Z$ in terms of those fields. Note, that the field $\mmm$ is auxiliary --- it is completely determined by the field $\nnn$. Indeed, the field $g$ is determined by $\nnn$ up to the U(1)$\times$U(1) phase factor. The columns $g_1(x),g_2(x),g_3(x)\in \mathbb{C}^3$ of the SU(3) matrix $g(x)$ are determined by the eigenvalue equations for the anti-hermitian matrix $\nnn(x)$:  $\nnn g_1=-\frac{\iu}{\sqrt{2}} g_1$, $\nnn g_2=\frac{\iu}{\sqrt{2}} g_2$, $g_3 = \overline{g_1}\times \overline{g_2}$ ($g^i_3=\epsilon^{ijk}\overline{g^j_1}\overline{g^j_2}$).

The forms $F$ and $G$ coincide with the Kirillov-Kostant symplectic forms on SU(3)/(U(1)$\times$U(1)) and SU(3)/(SU(2)$\times$U(1)) pulled-back with the maps $\nnn$ and $\mmm$:
$$
F=\frac{1}{4\sqrt{2}}(\nnn,[d\nnn,d\nnn]+[d\mmm,d\mmm]),\quad G=\frac{1}{4\sqrt{2}}(\mmm,[d\nnn,d\nnn]+[d\mmm,d\mmm]).
$$
We now calculate the form $Z=\nnn^* \overline{\sigma}$.
We decompose the form $\LLL$ into a form with values in the Cartan subalgebra of su(3) spanned by $\kappa_3$ and $\kappa_8$ and a form with values in a space orthogonal to this subalgebra:
$$
\LLL=\LLL_\perp+\LLL_\parallel.
$$
One can show, that (see e.g. \cite{FNK})
$$
g \LLL_\perp g^\dagger= \frac{1}{2} \left([\nnn,d\nnn]+[\mmm,d\mmm] \right)
$$
Using this formula we have:
\begin{eqnarray*}
\Tr{g \LLL_\perp g^\dagger \wedge d \left(g \LLL_\perp g^\dagger \right)}=\frac{1}{4} \Tr{\left([\nnn,d\nnn]+[\mmm,d\mmm] \right)\wedge \left([d\nnn,d\nnn]+[d\mmm,d\mmm] \right)}=\\=-\frac{1}{4}\left(\nnn, [d\nnn,[d\nnn,d\nnn]]+[d\nnn,[d\mmm,d\mmm]]\right)-\frac{1}{4}\left(\mmm, [d\mmm,[d\nnn,d\nnn]]+[d\mmm,[d\mmm,d\mmm]]\right),
\end{eqnarray*}
where $(C,D)=\Tr{C^\dagger D}$ is the SU(3) invariant scalar product in su(3), $[d\nnn,[d\nnn,d\nnn]]=d\nnn_a\wedge d\nnn_b\wedge d\nnn_c [\kappa_a,[\kappa_b,\kappa_c]]$ and similarly for other terms of this form. 
On the other hand:
\begin{eqnarray*}
\Tr{g \LLL_\perp g^\dagger \wedge d \left(g \LLL_\perp g^\dagger \right)}=\Tr{g\LLL_\perp g^\dagger\wedge g d \LLL_\perp g^\dagger}+\Tr{g\LLL_\perp g^\dagger\wedge dg g^\dagger \wedge g  \LLL_\perp g^\dagger}+ \Tr{g\LLL_\perp g^\dagger\wedge g  \LLL_\perp g^\dagger\wedge dg  g^\dagger}=\\=\Tr{\LLL_\perp \wedge d \LLL_\perp}+ 2\Tr{\LLL_\perp\wedge \LLL_\perp \wedge \LLL}=\Tr{\LLL_\perp \wedge d \LLL}+ 2\Tr{\LLL_\perp\wedge \LLL_\perp \wedge \LLL}=\Tr{\LLL_\perp\wedge\LLL_\perp\wedge \LLL_\perp}= 6\, \nnn^* \overline{\sigma}=6 Z.
\end{eqnarray*}
Therefore:
$$
Z=-\frac{1}{24}\left(\nnn, [d\nnn,[d\nnn,d\nnn]]+[d\nnn,[d\mmm,d\mmm]]\right)-\frac{1}{24}\left(\mmm, [d\mmm,[d\nnn,d\nnn]]+[d\mmm,[d\mmm,d\mmm]]\right).
$$
\section{Summary and discussion}
\subsection{Summary}
The maps $\nn:\mathbb{S}^3\to \SUU$ can be characterised by the third homotopy group of $\SUU$. We applied Novikov's procedure \cite{Novikov} to obtain the corresponding integral expression for this topological charge. The construction is as follows. We choose specific generators $[\overline{\kkt}]$ and $[\overline{\kke}]$ of the cohomology ring of $\SUU$. They are represented by closed forms $\overline{\kkt}$ and $\overline{\kke}$ given in equation \eqref{eq:basis}. The chosen generators satisfy $$[\overline{\kkt}]\wedge[\overline{\kkt}]+[\overline{\kke}]\wedge [\overline{\kke}]=0.$$ Therefore there exists a 3-form $\overline{\sigma}\in\Omega^3(\SUU)$ such that
$$
d\overline{\sigma}=\overline{\kkt}\wedge \overline{\kkt}+\overline{\kke}\wedge \overline{\kke}.
$$
Let $F,\,G,\,Z$ be pull-backs of the forms $\overline{\kkt},\, \overline{\kke},\, \overline{\sigma}$ with a map $\nn:\mathbb{S}^3\to \SUU$:
$$
F=\nn^* \overline{\kkt},\quad G=\nn^* \overline{\kke}, \quad Z=\nn^* \overline{\sigma}.
$$
Those pull-backs can be expressed in terms of sigma fields $\nnn=g \kappa_3 g^\dagger,\, \mmm=g \kappa_8 g^\dagger$, where $g:\mathbb{S}^3\to \rm{SU(3)}$. The forms $F$ and $G$ coincide with the pull-backs of the Kirillov-Kostant symplectic forms on $\SUU$ and SU(3)/(SU(2)$\times$U(1)) via the maps $\nnn$ and $\mmm$ respectively:
$$
F=\frac{1}{4\sqrt{2}}(\nnn,[d\nnn,d\nnn]+[d\mmm,d\mmm]),\quad G=\frac{1}{4\sqrt{2}}(\mmm,[d\nnn,d\nnn]+[d\mmm,d\mmm]).
$$
The form $Z$ expressed in terms of the fields $\nnn$ and $\mmm$ is:
$$
Z=-\frac{1}{24}\left(\nnn, [d\nnn,[d\nnn,d\nnn]]+[d\nnn,[d\mmm,d\mmm]]\right)-\frac{1}{24}\left(\mmm, [d\mmm,[d\nnn,d\nnn]]+[d\mmm,[d\mmm,d\mmm]]\right).
$$
Being pull-backs of closed forms, the forms $F$ and $G$ are also closed. Since $H^2_{\rm dR}(\mathbb{S}^3)=0$ there exist global one-forms $A,\, B\in \Omega^1(\mathbb{S}^3)$, such that:
$$
F=d A,\quad G=d B.
$$
Following Novikov \cite{Novikov} we define the topological charge to be
$$
Q[\nn]=\frac{1}{4\pi^2}\int_{\mathbb{S}^3} \left( A\wedge F + B\wedge G -Z \right).
$$
The charge differs from the Shabanov functional by a term $-\frac{1}{4\pi^2}\int_{\mathbb{S}^3}Z$. The charge $Q$ is homotopy invariant and takes only integer values --- in fact, it is an isomorphism of groups $\pi_3(\SUU)$ and $\mathbb{Z}$.
\subsection{Discussion}
Shabanov showed that the functional $\QS$ bounds from below the energy functional of the Faddeev-Niemi non-linear sigma model \cite{Faddeev_Niemi_SUN,Shabanov,FNK}:
$$
E[\nnn,\mmm]=E_2[\nnn,\mmm]+E_4[\nnn,\mmm],\ E_2[\nnn,\mmm]=\int_{\mathbb{R}^3} d^3x \left((\partial_i \nnn,\partial_i \nnn)+(\partial_i \mmm,\partial_i \mmm)\right),\ E_4[\nnn,\mmm]=\frac{1}{e^2}\int_{\mathbb{R}^3}d^3x\left(F_{ij}F_{ij}+G_{ij}G_{ij}\right).
$$ It would be very interesting to check, if the topological charge, that we obtained from the Novikov's procedure, still bounds from below the energy functional. Let us note, that the considerations from \cite{Shabanov} cannot be easily extended to the charge we study here. The bound found by Shabanov (as well as the Vakulenko-Kapitansky bound) strongly relies on an estimate of the form \cite{VK,Shabanov}:
$$
|\QS|\leq c (E_2 E_4)^{\frac{2}{3}}.
$$
Consider a map $U:\mathbb{S}^3\to \rm{SU(2)}$ of, say, degree 1, and the map $g_1:\mathbb{S}^3\to \rm{SU(3)}$ studied in the introduction: $(\go)\indices{^i_j}=\frac{1}{2}\Tr{\sigma_i U \sigma_j U^\dagger }$. A simple calculation shows that $E_4=0$, however $Q=4$. Therefore there will be no such estimate for the charge $Q$. On the other hand $E_2\not=0$ and $E_4=0$ means that the configuration of the sigma fields defined by $(\go)\indices{^i_j}=\frac{1}{2}\Tr{\sigma_i U \sigma_j U^\dagger }$ is not a stationary point of the energy functional (it follows from the virial theorem, see e.g. corollary 2.8 of \cite{JPhD}). Therefore a possible solution to this problem could be to look for an estimate for stationary points only.

Another important problem is to interpret this topological charge in the language of knotted strings. In the Faddeev-Skyrme model the sigma field $n$ is a map $n:\mathbb{S}^3\to \mathbb{S}^2$. A pre-image of a regular value $p\in\mathbb{S}^2$ is a closed curve. The Hopf integral can be interpreted as the linking number of pre-images $n^{-1}(p),n^{-1}(q)$ of any two different regular values $p,q\in \mathbb{S}^2$. Those pre-images are also lines of force of the magnetic field corresponding to the pull-back with the map $n$ of the Kirillov-Kostant symplectic form on $\mathbb{S}^2$. Interestingly, a pre-image of a point with the map $\nnn_{\frac{1}{2}}=\gh\kappa_3 \gh^\dagger$ is a closed curve (or empty set) whereas a pre-image of a point with the map $\nnn_1=\go\kappa_3 \go^\dagger$ is a set of points (or empty set). Therefore in the Faddeev-Niemi non-linear sigma model the topologically non-trivial excitations could be in principle either knot-like (as in the case of map $\nnn_{\frac{1}{2}}$) or point-like (as in the case of map $\nnn_1$) or could be a mixture of them.

It would be also of interest to extend the ideas to the $\rm{SU(N)}$ case (especially SU(4) since it may be important to non-static Euclidean SU(3) Yang-Mills theory --- see discussion section in \cite{FNK}). We expect that the calculation will be completely analogous to the calculation in the SU(3) case, a technical difficulty will be in finding the form $Z$ and expressing it in terms of sigma fields.
\section*{Acknowledgements}
The author gratefully acknowledges discussions with L. D. Faddeev. The author would like to thank for the hospitality he received at St. Petersburg Department of Steklov Mathematical Institute. This work was supported by the Foundation for Polish Science International PhD Projects Programme co-financed by the EU European Regional Development Fund.

\end{document}